\documentclass[12pt,preprint]{aastex}

\slugcomment{Nov. 21, 2006, V5}

\begin{document}

\title{Optical Spectroscopic Observations of CI Camelopardalis}

\author{Jingzhi Yan}\email{jzyan@pmo.ac.cn}
\author{Qingzhong Liu}\email{qzliu@pmo.ac.cn}
\author{Hengrong Hang}\email{hanghr@pmo.ac.cn}
 \affil{Purple Mountain Observatory,
Chinese Academy of Sciences, Nanjing, China\\}

\begin{abstract}
We present the results of optical spectroscopic observations of CI
Cam. Double-peaked profiles were simultaneously observed for the
first time in the hydrogen Balmer, He {\small I} $\lambda$6678 and
Fe {\small II} lines during an observational run in 2001 September.
An intermediate viewing angle of the circumstellar disk around the
B[e] star is consistent with our data. A significant decrease in the
intensity of the H$\alpha$ and He {\small I} lines in our 2004
September observations might have been the precursor of a line
outburst at the end of 2004. The remarkable increase in the
intensity of all lines and the decrease in visual brightness in 2005
might be due to the environment filling with new material ejected
during the outburst. The environment of CI Cam is influenced by mass
loss from the B[e] star and the outburst of its compact companion.
\end{abstract}

\keywords{binaries: close -- stars: emission-line, Be -- stars: individual CI Camelopardalis --
stars: spectroscopy - X-rays: binaries}

\section{INTRODUCTION}
\label{sect:intro} The bright transient X-ray source XTE J0421+560
was first observed by the ASM onboard the \emph{Rossi X-ray Timing
Explorer} ({\emph{RXTE}}) on 1998 March 31.64, and its peak X-ray
flux reached 2.0 crab on April 1.08 \citep{smith98}.
\citet{harmon04} also detected a flare in the 20-40 keV band with
BATSE during the outburst. Subsequently, the X-ray flux declined
rapidly, with an initial \emph{e}-folding time of $\sim$0.5 days for
the first 2 days, then slowing down to $\sim$2.3 days
\citep{belloni99}. Similar flux changes during the outburst were
also detected in other wavelength bands (optical, IR and radio) by
\citet{frontera98} and \citet{clark2000}.

\citet{hjellming98a} identified CI Cam (MWC 84) as the optical
counterpart to XTE J0421+560 just after its 1998 strong X-ray
outburst. Before the outburst, CI Cam was well studied in
photometric and spectroscopic observations
\citep{allen76,Downes1984,bergner95}. Since the outburst in 1998,
XTE J0421+560 has been an interesting transient X-ray source that
has been broadly studied in multiband (from radio to $\gamma$-ray)
observations \citep{hjellming98b,
belloni99,clark99,clark2000,orlandini00,robinson02,hynes02,miro02,boirin02,miodu04,ishida04}.
\citet{miro95} found an 11.7 day quasi period in the photometric
data of CI Cam, while \citet{barsukova05} found a 19.4-day orbital
period, based their 28-day long sets of observations. Variations
with the same period were also found in the radial velocities of He
{\small II} $\lambda$ 4686. However, the nature of the outburst is
not so clear due to the limited kinematical information about the
donor star. Different distances from 1 to $\geq$ 5 kpc based on
different methods have been found by different authors
\citep{zorec98,belloni99,clark2000,robinson02,hynes02,miro02,miodu04}.
With a closer distance of 1 kpc, \citet{orlandini00} suggested that
a thermonuclear runaway on a white dwarf trigged the 1998 outburst.
\citet{ishida04} also reached the same conclusion, that the compact
object in CI Cam is a white dwarf, based on the soft component of
the X-ray spectrum. However, according to the criteria for sgB[e]
stars \citep{zick86,lamers98} and XTE J0421+560's high X-ray
luminosity during the outburst, \citet{robinson02} and
\citet{hynes02} classified it as an sgB[e]/X-ray binary with a
neutron star or a black hole. \citet{simon06} explained the 1998
outburst as being due to thermal instability of the accretion disk,
analogous to the outburst of soft X-ray transients.

A high spatial resolution Very Large Array radio image of CI Cam
during outburst shows a relativistic cork-screw jet
\citep{hjellming98b}, just like the X-ray binary SS 433. But
\citet{rupen02} considered this result to be an artifact due to a
bad calibrator. \citet{miodu04} obtained radio imaging observations
during and after the 1998 outburst and suggested that CI Cam is a
shell-shocked X-ray nova analogous to an extragalactic supernovae
formed by the collapsar mechanism.

In this paper, we present the results of optical spectroscopic
observations of CI Cam from 2001 to 2005 and show its complicated
environment and evolution.

\section{OBSERVATIONS}

We obtained the spectra of CI Cam with the 2.16m telescope at the
National Astronomical Observatories, Chinese Academy of Sciences,
 from 2001 to 2005. The optical spectroscopy with an intermediate
resolution of 1.22 $\AA$ $pixel^{-1}$ was made with a CCD grating
spectrograph at the Cassegrain focus of the telescope. We took red
spectra covering from 5500 to 6700 $\AA$ and blue spectra covering
from 4300 to 5500 $\AA$ at different times. Sometimes low-resolution
spectra (covering from 4300 to 6700 $\AA$) were also obtained. The
journal of our observations is summarized in Table 1, including
observation date, Julian Date, wavelength range, exposure time, and
spectral resolution. All spectroscopic data were reduced with the
IRAF\footnote{IRAF is distributed by NOAO, which is operated by the
Association of Universities for Research in Astronomy, Inc., under
cooperation with the National Science Foundation.} package. They
were bias-subtracted and flat-field corrected, and had cosmic rays
removed. Helium-argon spectra were taken in order to obtain the
pixel-wavelength relations.

We select five typical spectra (covering the wavelength range from
6200 to 6650 $\AA$), which are plotted in Figure 1\emph{a}, in order
to see the evolution of the spectra of CI Cam from 2001 to 2005. All
spectra have been normalized with respect to the intensity of the
continuum. For comparison, we also plot the outburst spectrum of CI
Cam on 1998 April 18, adapted from \citet{hynes02}. The blue
spectrum (4400-5400 $\AA$) of CI Cam on 2002 October 27 is shown in
Fig. 1\emph{b}. The equivalent widths (EWs) of some selected lines
are listed in Table 1 and shown in Figure 2\emph{a}-2\emph{e}. The
uncertainty of the EWs is within 5\%, which mainly comes from the
level determination of the continuum. In order to see the EW
evolution of these lines since the 1998 outburst, we also plot the
data from \citet{barsukova02} and \citet{hynes02}, which are marked
with open symbols in Figure 2. The evolution of visual brightness
($V$) is shown in Figure 2\emph{f} and all the data in Figure
2\emph{f} are adopted from Barsukova et al.(2005). We analyze our
spectra and find the correlation between the variations of visual
brightness(\emph{V}) and our spectroscopic data.

\section{ANALYSIS AND RESULTS}
CI Cam has a rich emission-line spectrum with strong hydrogen
Balmer, He {\small I}, and numerous Fe {\small II} lines (Figure
1\emph{a}). The profiles of the hydrogen Balmer and He {\small I}
lines show a single peak during the time of quiescent state. During
the 1998 outburst, the strength of the Balmer and He {\small I}
lines dramatically increased to 2 times their pre-outburst levels,
and a weak He {\small II} $\lambda$4686 emission line could be seen
clearly \citep{wagner98}. All hydrogen Balmer lines returned to
their quiescent levels within about 30 days after the outburst,
while He {\small I} lines appeared to drop to a factor of 2--10
below their pre-outburst levels \citep{hynes02}. H$\alpha$ showed a
broad profile with a blue wing extending to at least - 2500 $ km\
s^{-1}$ and Fe {\small II} lines showed a symmetric and rectangular
profile with a flat top just after the 1998 outburst (Robinson et
al. 2002). \citet{hynes02} and \citet{miro02} obtained triple-peaked
Fe {\small II} emission lines in their high-resolution observations.
Forbidden lines such as [Fe {\small II}], [N {\small II}], [O
{\small I}], and [O {\small III}] were observed in the spectrum of
CI Cam. The presence of these forbidden lines indicates that the
circumstellar (CS) disk material is geometrically extended so that
there is a large amount of low-density gas (Lamers et al. 1998) in
the outer space around CI Cam.

Some new features are present in the spectra of CI Cam from our
observations from 2001 to 2005. We will discuss them in terms of
emission-line profile variability and EW evolution.

\subsection{The Profile Variability of Emission Lines}
Five characteristic H$\alpha$ profiles selected from each
observational run are presented in Figure 3\emph{a}. A double-peaked
profile of the H$\alpha$ emission line appears in our 2001 spectrum.
In the following observational runs, the H$\alpha$ line shows a
symmetrical single-peaked profile. The profile of the He {\small I}
$\lambda$6678 line follows the same evolutionary parth as the
H$\alpha$ line. Their full width at half-maximum (FWHM) is plotted
in Figure 4. We can see that H$\alpha$ and He {\small I}
$\lambda$6678 have a larger FWHM in our 2001 spectrum than they do
in 2005, while the FWHM of the He {\small I} $\lambda$5876 line,
with a single peak for all our spectra, has only a small variation
during our 5 years of observations.

Double-peaked emission profiles are simultaneously observed in the
hydrogen Balmer and He {\small I} $\lambda$6678 emission lines
during our 2001 observational run. Series of H$\alpha$ and He
{\small I} $\lambda$6678 lines taken in 2001 are plotted in Figure
3b, 3c, and the double-peaked $H\beta$ emission line detected on
2001 October 27 is seen in the inset of Figure 1b. To the best of
our knowledge, this is the first report of the double-peaked
profiles observed in the hydrogen Balmer and He {\small I}
$\lambda$6678 lines of CI Cam. Even when Fe {\small II} showed a
double-peaked or triple-peaked profile, the Balmer lines had a
single-peaked structure \citep{hynes02,miro02}. We can see an
asymmetric profile with a redshifted peak in the H$\alpha$ emission
line of 2001 September 25. In the spectrum of the next night, a
second shoulder appeared on the left side of the H$\alpha$ emission
line. Several minutes later, a clear double-peaked structure with
V/R$>$1 appeared in the second spectrum on 2001 September 26. The
typical profiles of the H$\alpha$ line on 2001 September 30 are
similar to those of the first line on September 26. For the last
spectrum in 2001, the H$\alpha$ emission line again showed a
double-peaked profile with V/R$>$1. Unlike the H$\alpha$ emission
line, the violet peak of He {\small I} $\lambda$6678 is always
higher than the red one (Figure 3c), except for the first spectrum
of 2001 September 25, in which the He {\small I} $\lambda$6678 line
only shows an asymmetric profile with a shoulder on the left side of
a redshifted single peak. The separation between the peaks or the
separation between the single peak and its shoulder kept nearly
constant for all the H$\alpha$ ($\Delta$$_p$=170 km $s^{-1}$) and He
{\small I} $\lambda$6678 ($\Delta$$_p$=209 km $s^{-1}$) lines during
our 2001 observations.

Many weak Fe {\small II} emission lines are also observed in the red
and blue arms of our spectra (Figure 1a and 1b). Here we only study
the uncontaminated Fe {\small II} $\lambda$6318 line in detail.
Figure 3(d) shows five Fe {\small II} emission lines selected from
each observational run. Just like H$\alpha$ and He {\small I}
$\lambda$6678, Fe {\small II} $\lambda$6318 in our 2001 spectra also
shows a double-peaked structure with a separation of 236 km
$s^{-1}$. Moreover, asymmetric Fe {\small II} profiles are also
observed during our 2002 and 2003 observations, while only single
symmetric lines are observed in our 2004 and 2005 spectra.

Forbidden lines also appear in the spectra of CI Cam. As an example,
we only study the forbidden lines of [N {\small II}] $\lambda$5755,
$\lambda$6548, and $\lambda$6583, due to the limited coverage of our
spectrum. The [N {\small II}] $\lambda$5755 line is unblended while
the [N {\small II}] $\lambda$6548, and $\lambda$6583 lines are in
the left and the right wings of the H$\alpha$ line, respectively.
The profile of the [N {\small II}] $\lambda$5755 line has a single
peak all the time, unlike the other lines described above. The [N
{\small II}] $\lambda$6548 line can be seen clearly only when the
broad left wing of the H$\alpha$ line becomes weak. The [N {\small
II}] $\lambda$6583 line is also mixed with the line of the C II
$\lambda$6583 line (A.S. Miroshnichenko 2005, private
communication). With the variability of the C II $\lambda$6583 line,
the [N {\small II}] $\lambda$6583 changes from a double-peaked
profile to a single-peaked one (Figure 1a).

\subsection{The EW Evolution of Selected Lines}

The EW evolution of selected lines is shown in Figure 2. The EWs of
H$\alpha$ and He {\small I} $\lambda$5876, $\lambda$6678 show a
similar temporal behavior, as shown in Fig. 2a, 2b, and 2c. Their
EWs dramatically increased during the 1998 outburst, then kept
declining on average. The EWs of H$\alpha$ and He {\small I} show a
remarkable decrease between our 2003 and 2004 observational runs,
especially for the He {\small I} $\lambda$5876 line, which had the
lowest emission level since the 1998 outburst.

The EW of Fe {\small II} increased dramatically during the 1998
outburst, just like the hydrogen Balmer and He {\small I} lines
(Fig. 2d). Subsequently, it declined to its pre-outburst level
quickly. During our observations, its relative intensity had a
subtle change on average, although it shows a fluctuation between
each night within each observational run.

During the 1998 outburst, the emission level of [N {\small II}]
$\lambda$5755 was relatively weak. After the outburst, its relative
intensity kept increasing and reached a peak in 1999 January. Since
then, the EW of [N {\small II}] $\lambda$5755 went through its
decline phase. Until 2004, [N {\small II}] $\lambda$5755 had a
steady emission level that was still above its pre-outburst level.

The EWs of the lines mentioned above, even those of [N {\small II}]
$\lambda$5755 and Fe {\small II} $\lambda$6318, had a rapid rise in
2005 October, especially the EWs of the He {\small I} $\lambda$5875
and He {\small I} $\lambda$6678 lines, which increased by a factor
of 2.

\section{DISCUSSION}
\subsection{The Properties of the CS Disk}
CI Cam is classified as a high mass X-ray binary, with a B[e]
supergiant donor star \citep{robinson02,hynes02} surrounded by dense
and absorbing CS material \citep{robinson02,boirin02,miodu04}. A CS
disk around a B[e] donor star is also suggested by \citet{hynes02}
and \citet{miro02}. There still exists a controversial debate on the
inclination of the CS disk. \citet{robinson02} suggested a
two-component wind model to explain their observations: a cool,
low-velocity, dense, and roughly spherical "iron wind" and a hot,
high-velocity wind. Based on the fact that the Fe {\small II}
emission profile changed from double-peaked to flat to double-peaked
during and after the outburst, \citet{hynes02} put forward a unified
wind model for CI Cam and concluded that most of the optical
emission lines, such as H Balmer, He {\small I} and Fe {\small II},
originate from a nearly pole-on CS disk. However, \citet{miro02}
suggested that the disk is inclined at an intermediate angle with
respect to the line of sight according to their triple-peaked Fe
{\small II} emission lines. For most of the time, the H Balmer and
He {\small I} lines of CI Cam showed a single-peaked profile. The
double-peaked profiles of the hydrogen Balmer, He {\small I}
$\lambda$6678, and Fe {\small II} lines simultaneously observed in
our 2001 spectra make us believe that the angle between the CS disk
and the line of sight is not so small (e.g., $\sim$30$^{\circ}$ ),
which coincides with the conclusion drawn by \citet{miro02}.

Double-peaked Fe {\small II} lines were observed by Hynes et al.
(2002) on 2000 December 1 (JD 2,451,879). Our double-peaked hydrogen
Balmer, He {\small I} $\lambda$6678 and Fe {\small II} lines were
observed between 2001 September 25 and 2001 October 1. At a later
time \citet{miro02} obtained triple-peaked Fe {\small II} lines in
high-resolution spectra on 2002 January 25 and February 4. In Figure
3d, asymmetric Fe {\small II} lines can still be seen clearly in the
spectra of 2002 and 2003. The variabilities of these double-peaked
lines show that great changes had taken place in the disk of the
B[e] star. A spherical wind is suggested during the 1998 outburst.
This situation has been discussed in detail by \citet{robinson02}
and \citet{hynes02}. The visual brightness, which is sensitive to
the density change in the CS disk, decreased after JD 2,451,200 and
reached a minimum around JD 2,452,141 (2001 August). The decline of
the visual brightness indicates that the density of the CS disk of
CI Cam was decreasing during that period, which might be due to the
dilution of the material ejected during the 1998 outburst. Moreover,
the central absorption in the H$\alpha$ line indicates that there
was not sufficient emitting matter with low radial velocities to
fill in the depression. Since the Fe II emission lines seem to have
formed in the inner regions of the CS disk (e.g., Hanuschik 1987),
they are the most likely to have double-peaked profiles.

\subsection{The Line Emission Outburst}

After the spectacular 1998 outburst, CI Cam rapidly went into its
quiescent state. There is not any obvious change on the \emph{RXTE}
ASM X-ray light curve from 2001 to 2005. Due to the limitation of
our observations, we could only get a glimpse at this exotic system
in each year. However, we still observed some peculiar phenomena
during our 5 year optical observations. Combining with the
photometric observations of CI Cam, we can trace the variability of
the environment of this peculiar source. A detailed study of the
physical parameters of the system is beyond the scope of this paper.
We only discuss some possible causes of the variability in the
optical spectroscopic and photometric observations.

The EWs of H$\alpha$ and He {\small I} $\lambda$$\lambda$5875, 6678
had an obvious decrease between our 2003 and 2004 observations,
while the visual brightness remained nearly unchanged during this
period (Fig. 2(f)). From Fig. 2(a-e) we can see that the EWs of the
He {\small I} lines had a steeper decrease than that of H$\alpha$,
and the EWs of Fe {\small II} $\lambda$6318 and [N {\small II}]
$\lambda$5755 remained constant on average between our 2003 and 2004
observations. Due to the limitation of our observational coverage,
the times when the decline began and ended could not be determined.
The duration of the decline was affected by the underlying physical
mechanism. If the drop in the emission strength was a short-term
event, it might have been due to a reduced recombination rate in the
disk. One possible trigger for this reduction would be a partial
loss of the stellar ionizing far-ultraviolet flux from the B[e]
star; another possibility would be that the highly increased matter
density (caused by ejecta, discussed below) affected the
recombination conditions by shielding the stellar radiation.

If a continuous decrease in the line strength occurred between our
2003 and 2004 observations, it would likely be due to the increasing
mass loss of the CS matter, which would result in a smaller CS disk.
Most materials might be dispersed into the environment of the
system, and a part of them would be accreted by the compact
companion. The increased accretion rate should enhance the X-ray
flux from the accretion disk. The absence of X-ray activity might be
due to X-ray absorption by the dense environment of CI Cam.

Another interesting phenomenon observed in 2005 is that the EWs of
all the lines rise, accompanied by a decrease in the visual
brightness (Fig. 2). The small increas in EW of the weak Fe {\small
II} and forbidden N {\small II} lines might have been caused by the
decrease of the continuum in the optical range. The anti-correlation
between emission level and brightness in Be stars has been discussed
by \citet{dachs82} and \citet{apparao91}. \emph{They suggested that
this kind of light variation is attributed to the absorption of the
central star light by the ejected material shell.} \citet{miro03}
and \citet{carciofi2006} also found an anti-correlation between the
EW of H$\alpha$ and visual brightness in the system of $\delta$
Scorpii. \citet{miro03} suggested that matter ejections into the
disk could strengthen the H$\alpha$ emission and cause an increase
of
 disk optical depth, consequently decreasing the visual brightness of the system.
\citet{carciofi2006} gave other possible reasons to explain this
erratic variability, such as star light being blocked by a warped CS
disk.

These scenarios can be applied to the system of CI Cam. One
possibility is that a shell of material was ejected from the B[e]
star as discussed above. The decrease of the line intensity in our
2004 observations might have been the precursor of the subsequent
line outburst, a phenomenon similar to that seen in {$\mu$ Centauri}
\citep{rivinius98}. The shell ejection would have resulted in a
global change of the mass distribution in the system. With the
expansion of the ejection, the ejected would become cool and
optically thick. The B[e] star might have been obscured by this cool
cloud in the line of sight, and its visual brightness began to
decline around JD 2,453,351 (2004 December). The observed increase
of the H$\alpha$ and He {\small I} EWs could have been due to an
enhancement of the density and volume of the line-emitting material
created by the ejection.

An alternative explanation, that large amount of matter was ejected
from the accretion disk around the compact object during an
outburst, cannot be ruled out with certainty. The physical mechanism
underlying the outburst remains under debate, although the favored
scenario at present is the disk instability mechanism (DIM; see
Lasota 2001 for a review). In the model of DIM, the outburst is
caused by a thermal-viscous instability cycle in the accretion disk.
Moreover, mass transfer variations are also an important element of
the DIM.

The ejection would have occurred around JD 2,453,351 (2004 December)
when the visual brightness began to decline (Fig. 2f). The outburst
might have been connected with the increasing accretion rate, which
might have been due to the increasing mass loss of the B[e] star or
an orbital phase change of the compact object. Lots of mass would
have accumulated in the accretion disk, and the DIM would have
resulted in an outburst. But the \emph{RXTE} ASM X-ray light curve
shows that no strong X-ray outburst was observed during the phase of
decrease of optical brightness that occurred in the end of 2004. The
2004 outburst might not be as strong as the 1998 outburst, and the
X-ray might have been smoothed by the dense environment of the B[e]
star \citep{miodu04}. The outburst could increase the density and
volume of the line emitting material. Thus, the EWs of H$\alpha$ and
He {\small I} showed an obvious increase after the burst. The X-ray
reprocessing into the envelope of the system might also made a
contribution to the EW increase during our 2005 observations.

The ejection at the end of 2004 might also have changed the
environment of CI Cam. After about 1 year's expansion, the emitting
CS disk became more extended, and the smallest FWHM of H$\alpha$ was
obtained in our 2005 observations. When the ejected materials were
dispersed into the outer environment of the system, the visual
brightness of CI Cam began to increase around 2006 January (see in
the website of E.A. Barsukova).

\section{Conclusions}

We present and analyze our optical spectroscopic observations of CI
Cam from 2001 to 2005. Combining with the evolution of the visual
brightness, we make the following major findings about the system of
CI Cam: \flushleft
\begin{enumerate}

\item The double-peaked profiles of optically thick hydrogen
Balmer, He {\small I} $\lambda$6678 and Fe {\small II} lines in
our 2001 spectra indicate a CS disk with an intermediate
inclination angle.

\item Great changes have taken place in the environment of CI Cam since its 1998 strong outburst.
The decrease in visual brightness and the disappearance of the
central narrow emission component in the H$\alpha$ line in 2001
indicate that matter from the outer parts of the disk (which has the
lowest radial velocities) has left the system. We interpret the
decrease in the intensity of H$\alpha$ and He {\small I} lines in
our 2004 observations as being due the reduction of the
recombination rate in the circumstellar disk. Two possibilities are
suggested for the reduction of the recombination rate. We also
consider another scenario for the decrease of line intensity that
the mass loss might result in a smaller circumstellar disk.

\item The EW decrease of the H$\alpha$ and He {\small I} lines in our 2004 observations might have been the
precursor of the line outburst at the end of 2004. Much of the
material has been ejected into the environment of CI Cam during the
outburst. The B[e] star might be obscured by the ejected clouds of
gas, and the visual brightness began to decrease around JD 2,453,351
(2004 December). Scenarios of ejection from the B[e] star and from
the accretion disk are both considered. With the expansion of the
ejecta, stronger and narrower emission lines were obtained during
our 2005 observations.

\end{enumerate}

The environment of CI Cam is influenced by the variabilities of the
B[e] star and its compact companion. In order to unveil the secrets
of this peculiar object, frequent spectroscopic and photometric
observations are needed.

\acknowledgements

We are grateful to the anonymous referee for useful comments and
suggestions, to A. S. Miroshnichenko and W. Hummel for useful
discussions and suggestions, and to R. I. Hynes and E. A. Barsukova
for access to their observational data. This research is partially
supported by the National Natural Science Foundation of China under
grants 10433030 and 10673032.

\begin{figure*}

\includegraphics[bb=15 15 290 215,width=8cm]{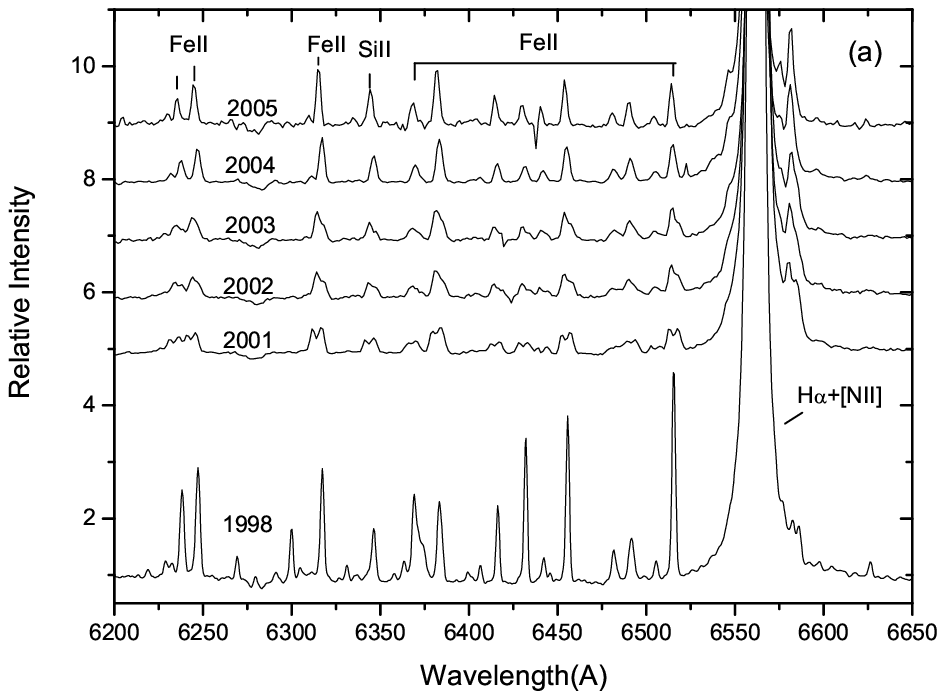}%
\includegraphics[bb=15 15 290 215,width=8cm]{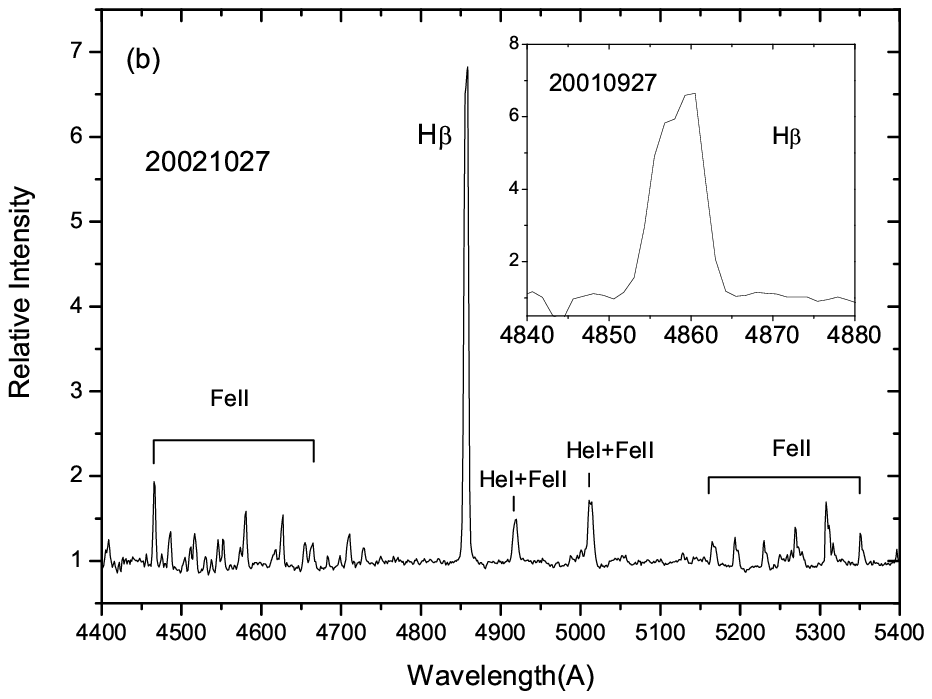}

\caption{Optical spectroscopic observations of CI Cam from 2001 to
2004. \textbf{a}) Red spectra. The spectrum of 1998 is adapted from
Hynes et al. (2002).  \textbf{b}) Blue spectrum. The inset is the
profile of $H\beta$ on 2001 September 27. }
\end{figure*}

\begin{center}

\begin{figure*}
\centering
\includegraphics[bb=15 15 290 215,width=8cm]{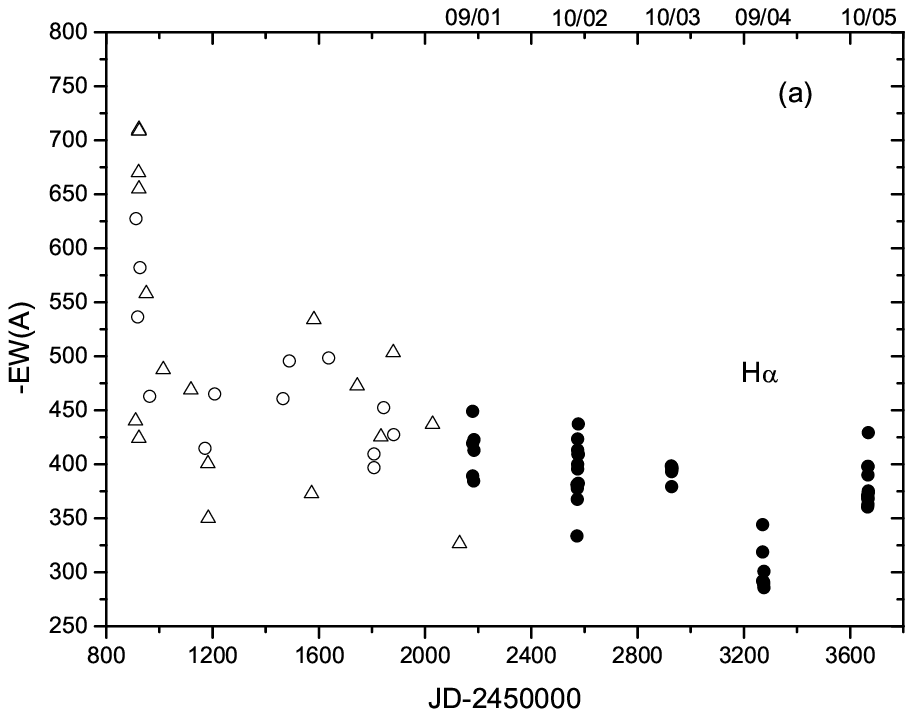}%
\includegraphics[bb=15 15 290 215,width=8cm]{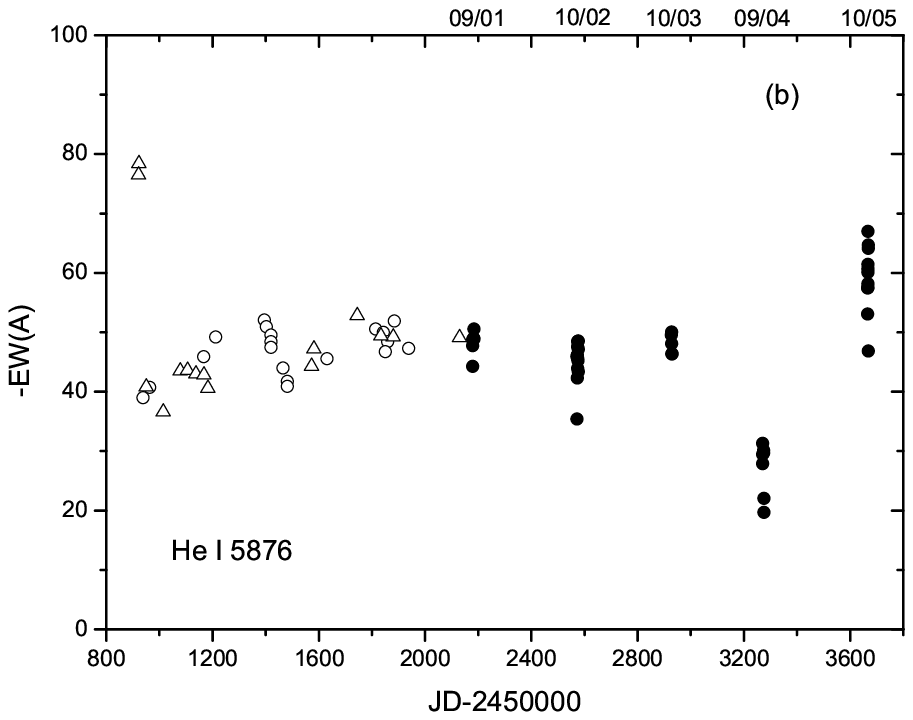}
\includegraphics[bb=10 15 285 215,width=8cm]{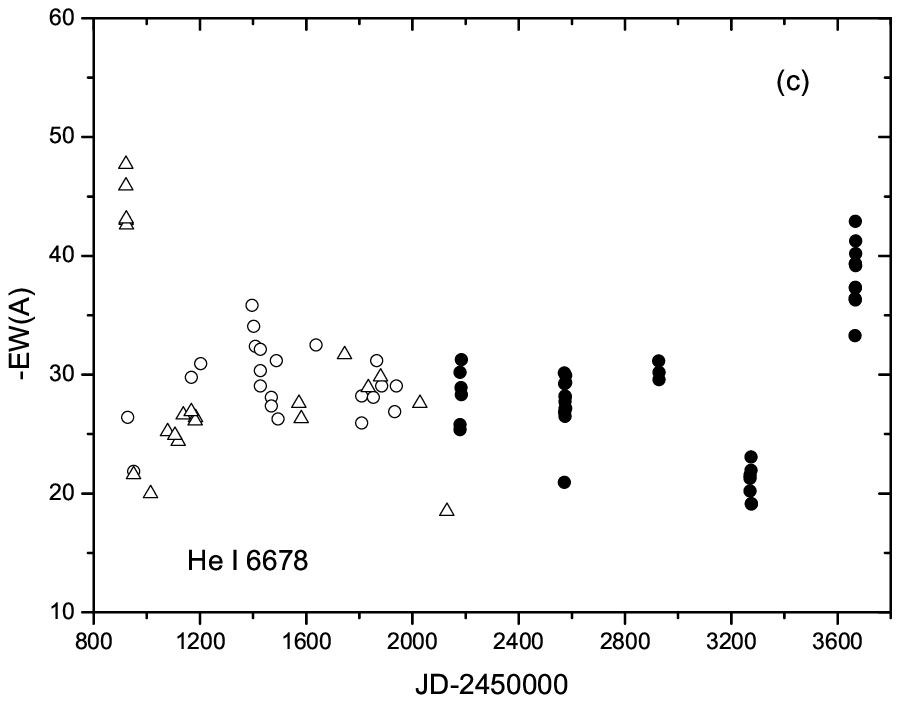}%
\includegraphics[bb=10 15 285 215,width=8cm]{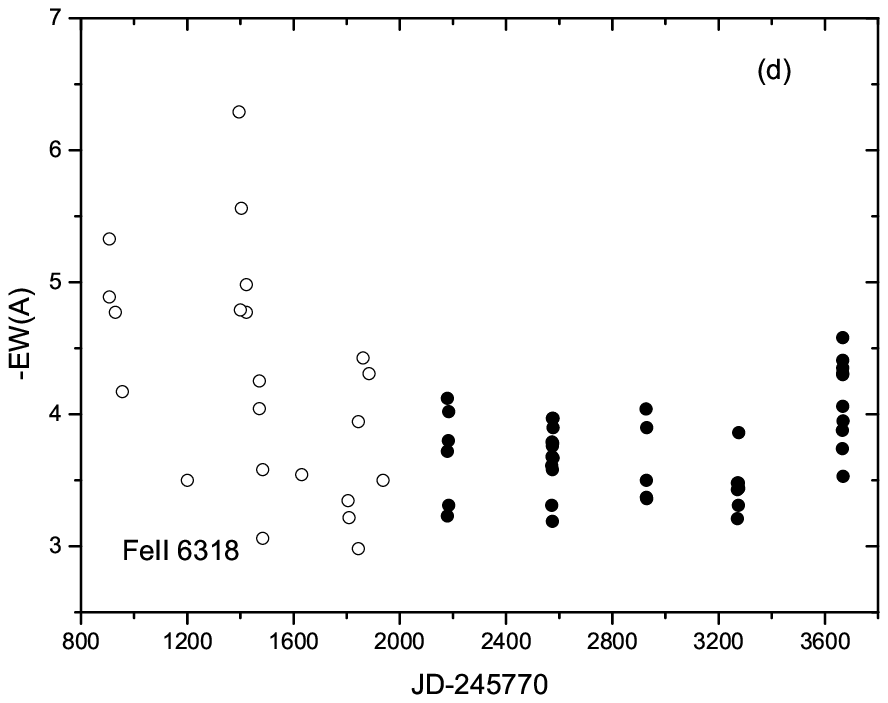}
\includegraphics[bb=10 15 283 215,width=8cm]{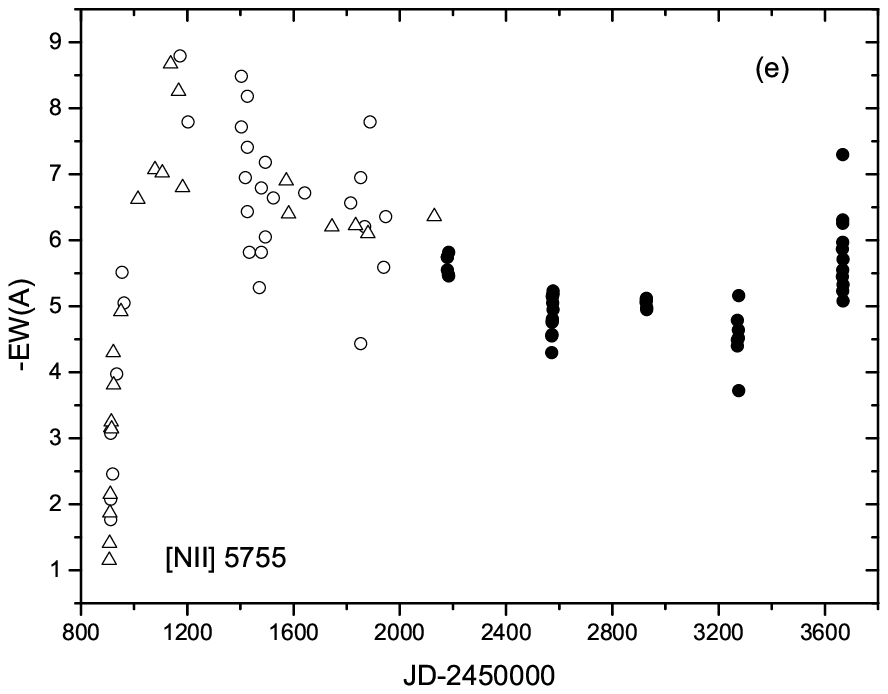}%
\includegraphics[bb=20 15 290 215,width=8cm]{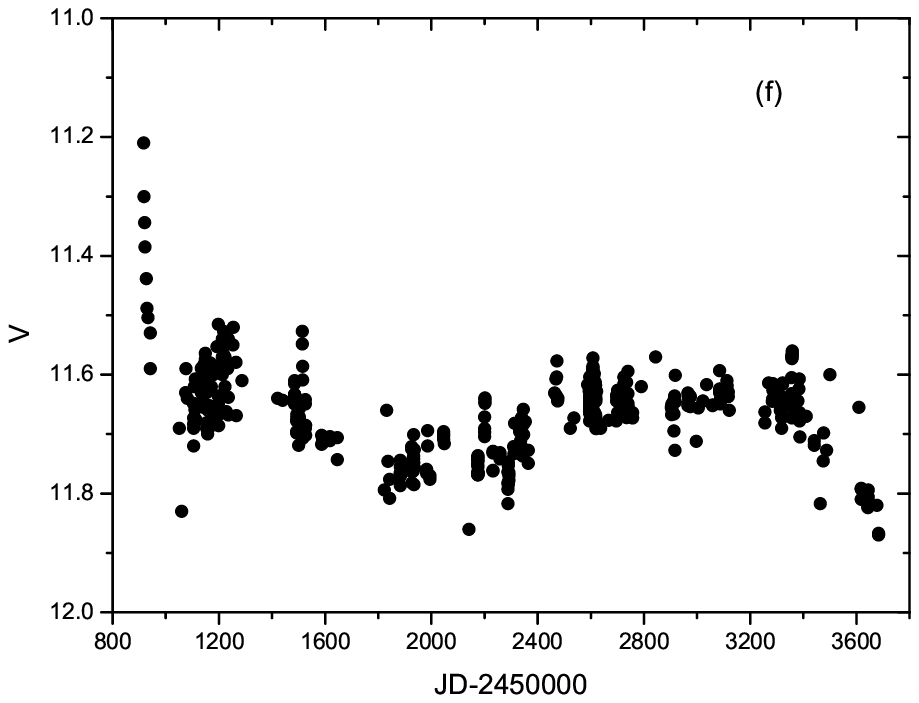}

\caption{\textbf{a-e}) EW evolution of some selected lines. Because
of the difficulty in determining the level of continuum and
contamination by other lines, the statistical error is not given in
the chart. In order to see the evolution clearly, we adapt the data
after 1998 outburst from Barsukova et al. (2002; circle) and Hynes
et al. (2002; triangle) and plot them with open symbols. \textbf{f})
Evolution of visual brightness from Barsukova E. A. et al.(2005).}

\end{figure*}
\end{center}

\begin{center}
\begin{figure*}
\centering
\includegraphics[bb=15 15 290 215,width=8cm]{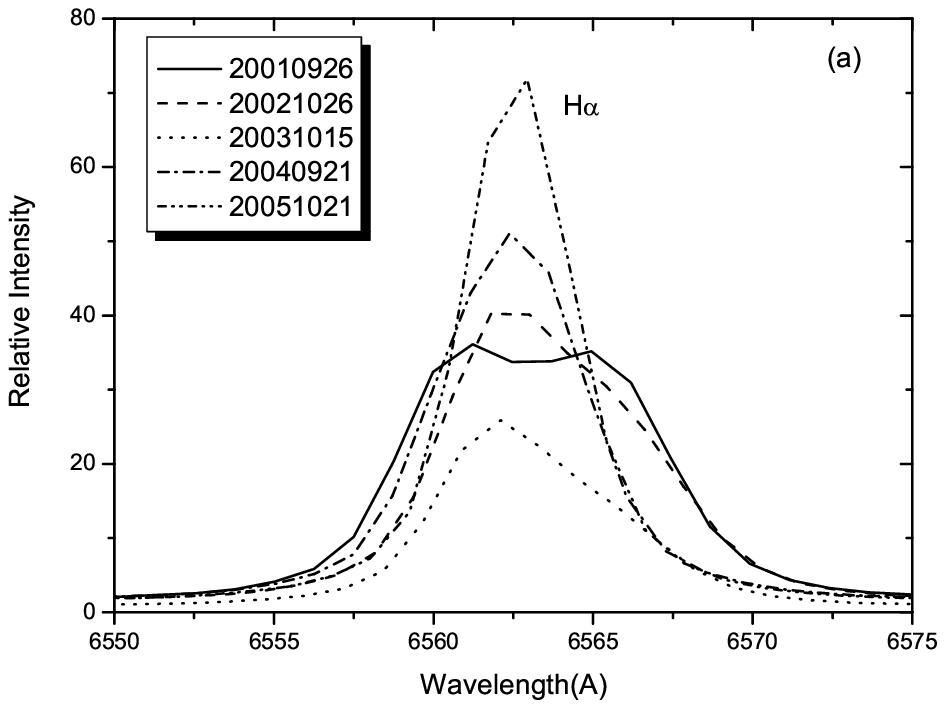}%
\includegraphics[bb=15 15 290 215,width=8cm]{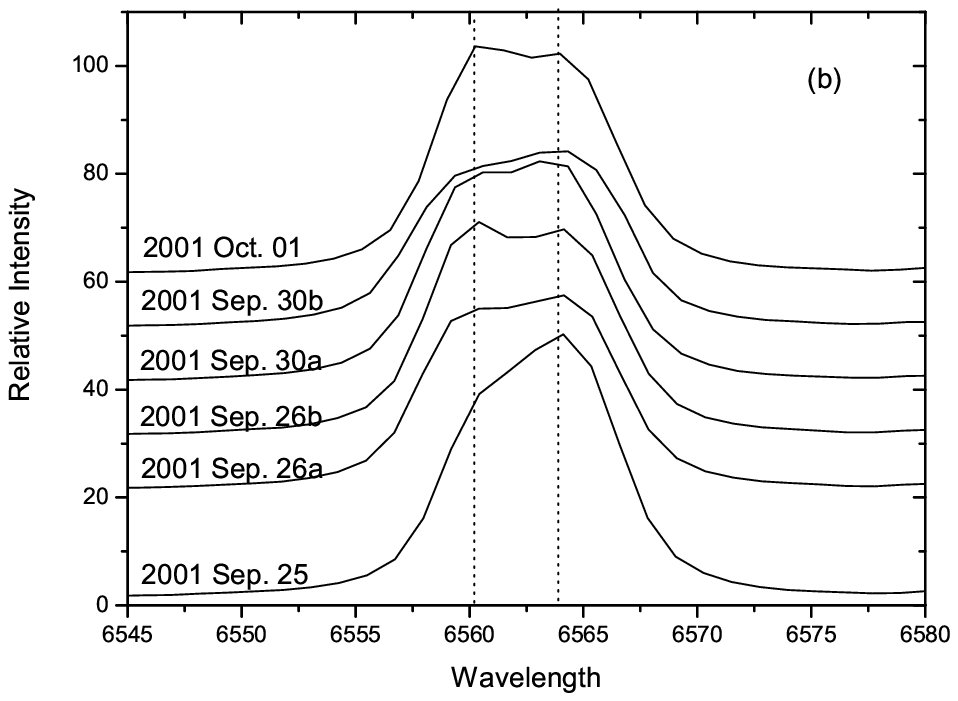}
\includegraphics[bb=15 15 290 215,width=8cm]{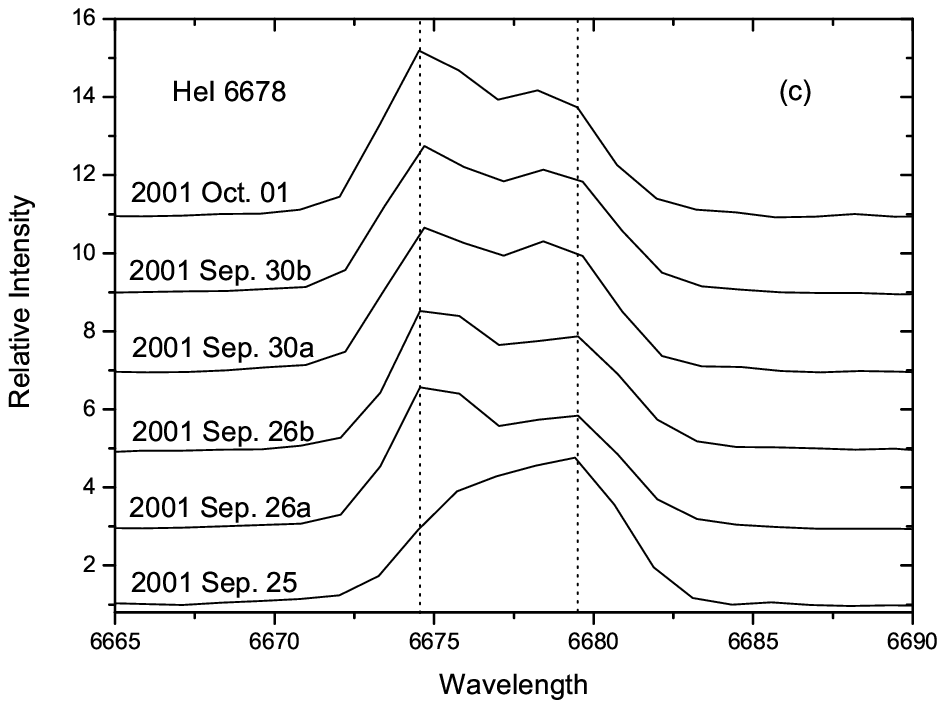}%
\includegraphics[bb=10 15 280 215,width=8cm]{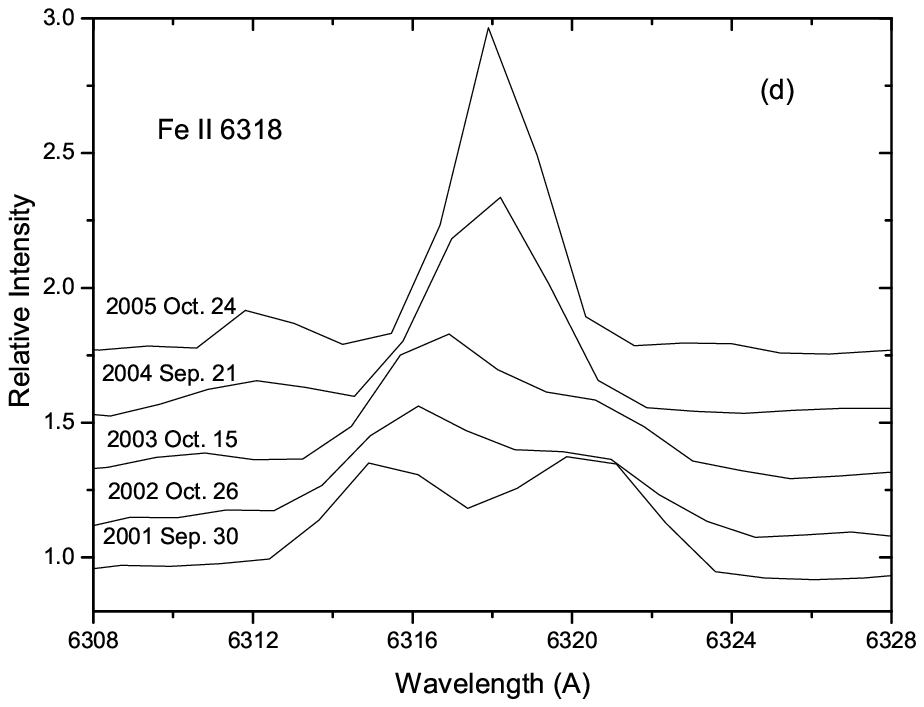}
\caption{\textbf{a}) Selected H$\alpha$ emission lines in our
observations. \textbf{b})$H\alpha$ emission lines in our 2001
observations. \textbf{c}) He {\small I} $\lambda$6678 in our 2001
observations. \textbf{d}) Fe {\small II} $\lambda$6318 profiles
during our 5 years of observations}
\end{figure*}
\end{center}

\begin{center}

\begin{figure*}
\centering
\includegraphics[bb=100 140 450 800, width=8cm]{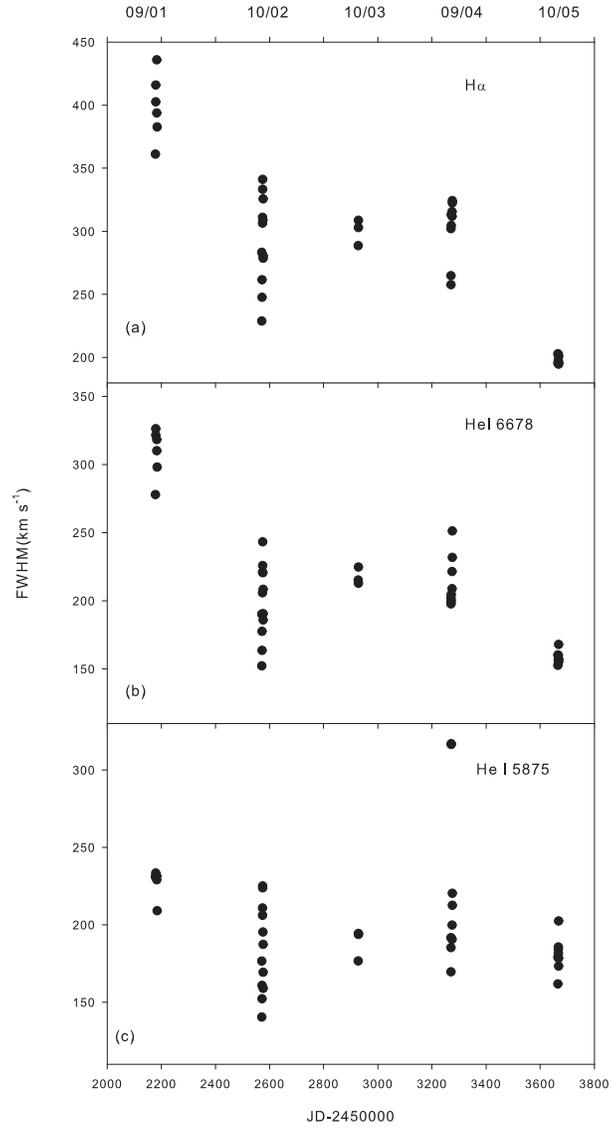}

\caption{FWHM evolutions of the $H\alpha$ and He {\small I} lines of
CI Cam from 2001 to 2005}
\end{figure*}
\end{center}

\clearpage
\begin{deluxetable}{lccccccccc}
\tablecolumns{9} \tablecaption{Summary of the spectroscopic observations of CI Cam.}
\tablewidth{0pc} \tablehead{\colhead{1}&\colhead{2} & \colhead{3} & \colhead{4}  & \colhead{5} &
\colhead{6} &\colhead{7} &\colhead{8} &\colhead{9} &\colhead{10}} \startdata
20010925  &2452178.303   &5550-6750 &600  &1.2   & 449.02 & 44.30 &25.37 &4.12 &5.75   \\
20010926  &...2179.300   &5550-6750 &600  &1.2   & 389.17 & 48.75 &25.81 &3.72 &5.55   \\
20010926  &...2179.305   &5550-6750 &200  &1.2   & 419.31 & 47.77 &30.21 &3.23 &5.74   \\
20010927  &...2180.308   &4300-5500 &600  &1.2   &-       &-      &-     &-    &-      \\
20010930  &...2183.308   &5550-6750 &600  &1.2   & 384.67 & 49.02 &28.91 &3.80 &5.48   \\
20010930  &...2183.313   &5550-6750 &200  &1.2   & 412.88 & 50.56 &28.31 &3.31 &5.82   \\
20011001  &...2184.316   &5550-6750 &200  &1.2   & 422.86 & 48.85 &31.27 &4.02 &5.46   \\
20021023  &...2571.262   &5550-6750 &500  &1.2   & 333.61 & 35.43 &20.93 &3.61 &4.55   \\
20021023  &...2571.266   &5550-6750 &100  &1.2   & 381.18 & 45.92 &30.15 &3.31 &4.30   \\
20021024  &...2572.301   &5550-6750 &400  &1.2   & 367.72 & 42.31 &26.83 &3.68 &4.57   \\
20021024  &...2572.305   &5550-6750 &200  &1.2   & 377.83 & 46.17 &29.25 &3.79 &4.76   \\
20021026  &...2574.236   &5550-6750 &500  &1.2   & 413.19 & 43.96 &27.15 &3.58 &4.81   \\
20021026  &...2574.241   &5550-6750 &200  &1.2   & 423.44 & 48.48 &26.51 &3.97 &5.16   \\
20021026  &...2574.328   &5550-6750 &200  &1.2   & 400.01 & 45.70 &27.74 &3.19 &4.78   \\
20021026  &...2574.333   &5550-6750 &500  &1.2   & 395.61 & 47.60 &28.20 &3.76 &5.14   \\
20021027  &...2575.150   &4300-5500 &300  &1.2   &-       &-      &-     &-    &-      \\
20021027  &...2575.200   &4300-5500 &1000 &1.2   &-       &-      &-     &-    &-      \\
20021027  &...2575.278   &5550-6750 &400  &1.2   & 409.26 & 45.23 &28.12 &3.78 &5.05   \\
20021028  &...2576.379   &5550-6750 &100  &1.2   & 409.33 & 47.12 &29.33 &3.67 &5.18   \\
20021028  &...2576.270   &5550-6750 &300  &1.2   & 382.38 & 43.39 &27.16 &3.97 &5.23   \\
20021028  &...2576.275   &5550-6750 &200  &1.2   & 437.21 & 48.55 &29.96 &3.90 &4.95   \\
20031014  &...2927.285   &5550-6750 &400  &1.2   & 398.34 & 49.57 &31.15 &4.04 &5.06   \\
20031015  &...2928.333   &5550-6750 &300  &1.2   & 393.13 & 48.11 &29.59 &3.37 &5.10   \\
20031015  &...2928.335   &5550-6750 &60   &1.2   & 379.38 & 50.08 &30.20 &3.50 &5.12   \\
20031016  &...2929.224   &4300-6700 &50   &2.4   & 395.32 & 46.36 &-     &3.90 &4.95   \\
20031016  &...2929.226   &4300-6700 &50   &2.4   & 397.33 & 46.42 &-     &3.36 &4.98   \\
20040921  &...3270.311   &5550-6750 &1500 &1.2   & 291.88 & 27.87 &20.22 &3.43 &4.49   \\
20040921  &...3270.326   &5550-6750 &900  &1.2   & 318.73 & 31.34 &21.27 &3.48 &4.79   \\
20040921  &...3270.334   &5550-6750 &300  &1.2   & 344.19 & 29.48 &21.54 &3.21 &4.40   \\
20040925  &...3274.323   &5550-6750 &800  &1.2   & 289.26 & 30.12 &23.08 &3.48 &4.64   \\
20040925  &...3274.330   &5550-6750 &300  &1.2   & 290.90 & 29.69 &21.95 &3.31 &4.52   \\
20040926  &...3275.293   &5550-6750 &300  &1.2   & 285.85 & 22.07 &19.14 &3.44 &5.16   \\
20040926  &...3275.297   &5550-6750 &200  &1.2   & 300.97 & 19.72 &19.12 &3.86 &3.72   \\
20051021  &...3665.348   &5550-6750 &400  &1.2   & 360.54 & 53.09 &33.29 &3.74 &5.87   \\
20051021  &...3665.352   &5550-6750 &200  &1.2   & 370.84 & 57.49 &36.39 &3.88 &5.45   \\
20051023  &...3667.232   &4300-6700 &50   &2.4   & 368.01 & 58.21 &36.3  &4.41 &5.97   \\
20051023  &...3667.234   &4300-6700 &30   &2.4   & 397.80 & 61.46 &39.34 &4.35 &5.55   \\
20051023  &...3667.235   &4300-6700 &100  &2.4   & 363.11 & 67.01 &37.35 &4.58 &6.26   \\
20051023  &...3667.342   &5550-6750 &200  &1.2   & 390.25 & 60.11 &42.92 &4.30 &5.23   \\
20051023  &...3667.344   &5550-6750 &100  &1.2   & 398.08 & 60.72 &39.33 &4.06 &6.31   \\
20051023  &...3667.345   &5550-6750 &60   &1.2   & 368.82 & 57.52 &37.30 &4.31 &7.30   \\
20051024  &...3668.290   &5550-6750 &200  &1.2   & 375.34 & 46.83 &39.19 &3.53 &5.08   \\
20051024  &...3668.292   &5550-6750 &100  &1.2   & 373.91 & 64.16 &41.26 &3.95 &5.71   \\
20051024  &...3668.294   &5550-6750 &100  &1.2   & 429.34 & 64.69 &40.18 &-    &5.33   \\
20051027  &...3671.296   &4300-5500 &200  &1.2   &-       &-      &-     &-    &-      \\
20051027  &...3671.305   &4300-5500 &1000 &1.2   &-       &-      &-     &-    &-      \\
\enddata
\tablenotetext{Note} {\textbf{Col. 1}: Date(yyyymmdd); \textbf{Col. 2}: Julian Date; \textbf{Col.
3}: Wavelength Range(\AA); \textbf{Col. 4}: Exposure Time(s); \textbf{Col. 5}: Spectral
Resolution(\AA/pix); \textbf{Col. 6}: The Equivalent Width of $H\alpha$ (-\AA); \textbf{Col. 7}: EW
of He {\small I} $\lambda$ 5875(-\AA); \textbf{Col. 8}: EW of He {\small I} $\lambda$ 6678(-\AA);
\textbf{Col. 9}: EW of Fe {\small II}
$\lambda$ 6318(-\AA); \textbf{Col. 10}: EW of [N {\small II}] $\lambda$ 5755(-\AA).\\
Because of the difficulty in determining the level of continuum and
contamination by other lines, the statistical errors are not given
in the table.}

\end{deluxetable}

\end{document}